\documentclass[12pt]{article}
\usepackage{graphicx}

\newcommand\pubnumber{CIPANP2018-Franklin}
\newcommand\pubdate{\today}

\def\cmu{Department of Physics\\
Carnegie Mellon University, Pittsburgh, PA, 15213}
\def\support{\footnote{Primary support for US participation in KATRIN is provided by the U.S. Department
of Energy Office of Science, Office of Nuclear Physics, under award number
DE-FG02-97ER41020.}}

\def\Title#1{\begin{center} {\Large #1 } \end{center}}
\def\Author#1{\begin{center}{ \sc #1} \end{center}}
\def\Address#1{\begin{center}{ \it #1} \end{center}}

\newcommand\pubblock{\rightline{\begin{tabular}{l} \pubnumber\\
         \pubdate  \end{tabular}}}
\newenvironment{Abstract}{\begin{quotation}  }{\end{quotation}}
\newenvironment{Presented}{\begin{quotation} \begin{center} 
             PRESENTED AT\end{center}\bigskip 
      \begin{center}\begin{large}}{\end{large}\end{center} \end{quotation}}
\def\Acknowledgements{\bigskip  \bigskip \begin{center} \begin{large}
             \bf ACKNOWLEDGEMENTS \end{large}\end{center}}

\def\beq{\begin{equation}}
\def\eeq#1{\label{#1}\end{equation}}
\def\eeqn{\end{equation}}

\def\beqa{\begin{eqnarray}}
\def\eeqa#1{\label{#1}\end{eqnarray}}
\def\eeqan{\end{eqnarray}}

\let\bar=\overbar

\def\Dslash{\not{\hbox{\kern-4pt $D$}}}
\def\dslash{\not{\hbox{\kern-2pt $\del$}}}

\def\msb{{\bar{\ssstyle M \kern -1pt S}}}

\begin{document}
\begin{titlepage}
\pubblock

\vfill
\Title{The KATRIN Neutrino Mass Measurement: Experiment, Status, and Outlook}
\vfill
\Author{ G. B. Franklin  for the KATRIN Collaboration\support}
\Address{\cmu}
\vfill
\begin{Abstract}
The Karlsruhe Tritium Neutrino (KATRIN) experiment will provide a measurement of the effective electron-neutrino mass, $m(\nu_e)$, based on a precision measurement of the tritium beta decay spectrum near its endpoint.  The effective mass is an average of the neutrino mass eigenvalues $m_i$ weighted by the flavor-mass mixing parameters $U_{ei}$ according to the relation $m^2(\nu_e)= \sum_{i=1}^3 |U_{ei} |^2 m_i^2$.  The KATRIN apparatus uses a windowless gaseous tritium source (WGTS) and a spectrometer based on the MAC-E filter concept to measure the beta energy spectrum.  The KATRIN program is designed to reach a mass sensitivity of 0.2~eV (90\% C.L.).  The collaboration has completed a series of commissioning measurements and is moving into the first running of tritium.  The KATRIN measurement technique, early commissioning results,
and the future outlook will be presented.
\end{Abstract}
\vfill
\begin{Presented}
The Conference on the Intersections of Particle and Nuclear Physics\\
Palm Springs, CA,  May 29 - June 3, 2018
\end{Presented}
\vfill
\end{titlepage}
\def\thefootnote{\fnsymbol{footnote}}
\setcounter{footnote}{0}

\section{Introduction}
While the bulk of our knowledge of neutrino masses comes from oscillation measurements, oscillation experiments are sensitive to the differences of the squares of the masses
and do not determine the absolute neutrino mass scale.

There are several methods for getting at the absolute value of the masses.   For example, the spread in arrival time of neutrinos from supernova events has provided insight.
Observations of neutrinoless double beta decay could potentially provide information.   
Neutrino masses may also be observed via kinematic effects; for example, the shape of a beta-decay electron spectrum is sensitive
to the neutrino masses.
At this time, the absolute mass scale has not been determined and the best upper limits on the neutrino masses from direct measurements
give an upper limit of the effective electron-neutrino mass of $m_{\nu_e}<2$ eV (90\% C.L)\cite{Kraus,Aseev,Olive}.    Lower limits have been inferred from the structure in the Cosmic Microwave Background.\cite{Couchot}
The KATRIN experiment uses a high precision measurement of the tritium beta decay spectrum for a model-independent determination  of $m_{\nu_e}$.
\begin{figure}[htb]
\centering
\includegraphics[height=1.5in]{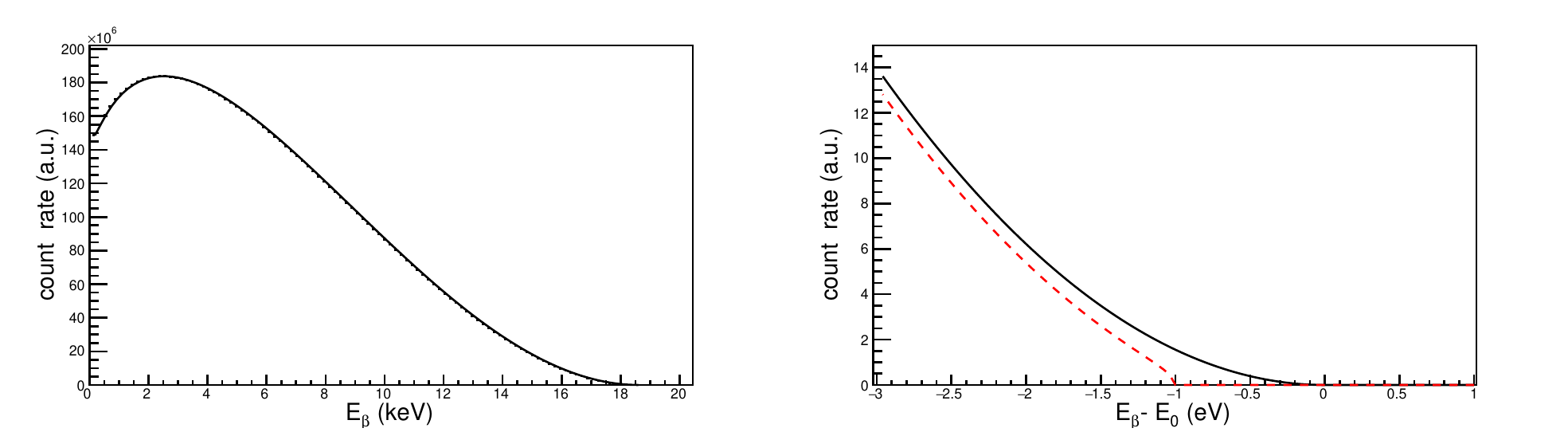}
\caption{(Left) The $^3\rm{H}$ beta-decay spectrum.  (Right) The  spectrum end-point calculated for two effective neutrino masses, $m_{\nu_e}=0$ (solid black) and $m_{\nu_e}=1$~eV. (dashed red)}
\label{fig:decayDiagram}
\end{figure}

The calculated energy spectrum of the electrons emitted from the triton beta-decay,
$^3$H$\rightarrow ^3$He$+e+\bar{\nu}_e$, is shown in Fig.~\ref{fig:decayDiagram}a.  In principle, the beta-energy endpoint, or maximum electron energy, is determined by the lightest neutrino mass that couples to the beta decay process.  The measured spectrum would then have kinks at slightly lower energies corresponding to the kinematic endpoints for each neutrino mass state.   However, assuming mass separations much smaller than the experimental energy resolution, the spectrum is only sensitive to an effective electron-neutrino mass,  $m_{\nu_e}$ defined by averaging the squares of the masses weighted by the mass-eigenvalues to electron-neutrino mixing amplitudes,  $U_{ej}$,
$$
m_{\nu_e} \equiv \sum_j |U_{ej}|^2  m_j^2.
$$
Figure~\ref{fig:decayDiagram}b shows an expanded view of the spectrum
near its endpoint as calculated for two different effective electron neutrino masses,  $m_{\nu_e}=0$ and $m_{\nu_e}=1$~eV.  It can be seen that the 
spectrum for  $m_{\nu_e}=1$~eV is shifted relative to the  $m_{\nu_e}=0$ and exhibits an altered curvature near its endpoint.  Thus a precision
measurement of the spectrum shape has the potential to extract the effective electron-neutrino mass, $m_{\nu_e}$.

\section{The KATRIN Apparatus}
The Karlsruhe Tritium Neutrino experiment, or KATRIN, is sensitive to an effective neutrino mass down to about
 0.2 ~eV.\cite{Angrik:2004,Amsbaugh:2015}  It has two major components:  an intense tritium source of beta decay particles and a high resolution integrating spectrometer with better than 1~eV resolution.
The electrons from the $^3$H$\rightarrow ^3$He$+e+\bar{\nu}_e$  decay are detected with a segmented silicon detector,
 but the eV scale energy resolution is provided by a MAC-E spectrometer.  The shape of the spectrum near its endpoint will be used to determine
 the value of $m_{\nu_e}$.
 
\begin{figure}[htb]
\centering
\includegraphics[height=1.2in]{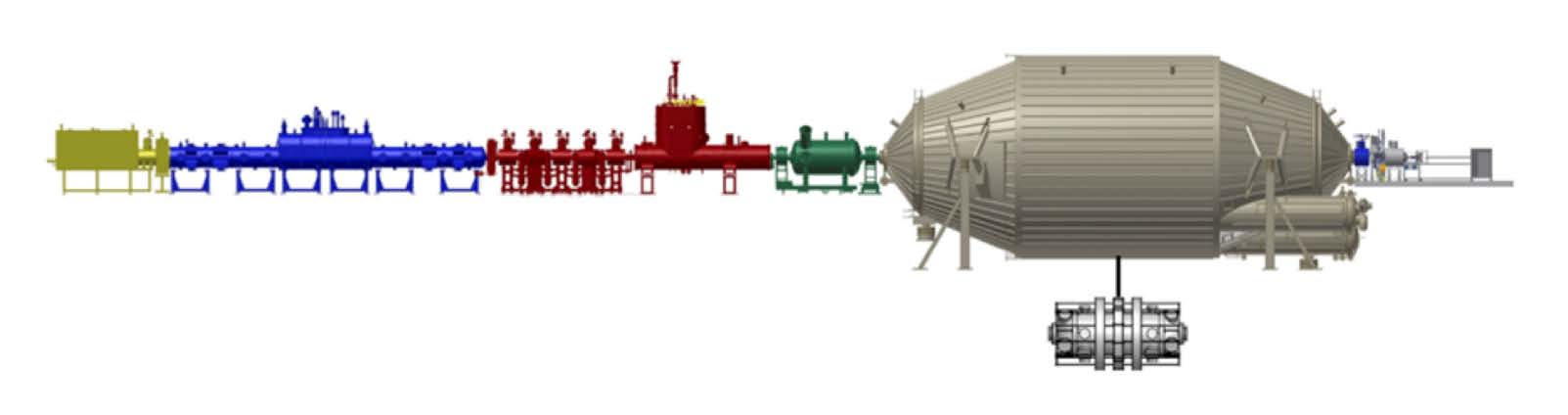}
\caption{The KATRIN apparatus.  The major components are, listing from left to right, the Rear Wall (yellow), the Windowless Gaseous Tritium Source (blue),
the differential and cryogenic pumping sections (red),  the Pre-Spectrometer (green), the Main Spectrometer (gray), and the section housing the forward focal plane detector section shown on the right.}
\label{fig:KATRIN}
\end{figure}

The layout of the 70 meter long KATRIN apparatus is shown in Fig.~\ref{fig:KATRIN}.
On the left,  in blue, is the Windowless Gaseous Tritium Source (WGTS).   Tritium is pumped in at around $10^{-3}$ mbar, providing a source of $10^{11}$ beta decays per second.  The beta-decay electrons are transported from the left to the right in the figure; they originate in the WTGS
and follow magnetic field lines of a few Tesla through the Main Spectrometer (gray) and then on to the Focal Plane Detector (FPD) located to the right of
the Main Spectrometer.
 
The required sub-eV resolution prohibits the use of a window between the gaseous tritium source and the spectrometer.   In lieu of a window at
the exit of the WGTS,  a series of differential and cryogenic pumping stations  reduce the tritium concentration by 14 orders of magnitude, maintaining an ultra-high vacuum in the main spectrometer vessel that is essentially tritium-free.   The Pre-Spectrometer, located just before the main spectrometer, is set to block all the electrons except the few percent near the beta-spectrum endpoint.

\section{The MAC-E Spectrometer}

\begin{figure}[htb]
\centering
\includegraphics[height=1.2in]{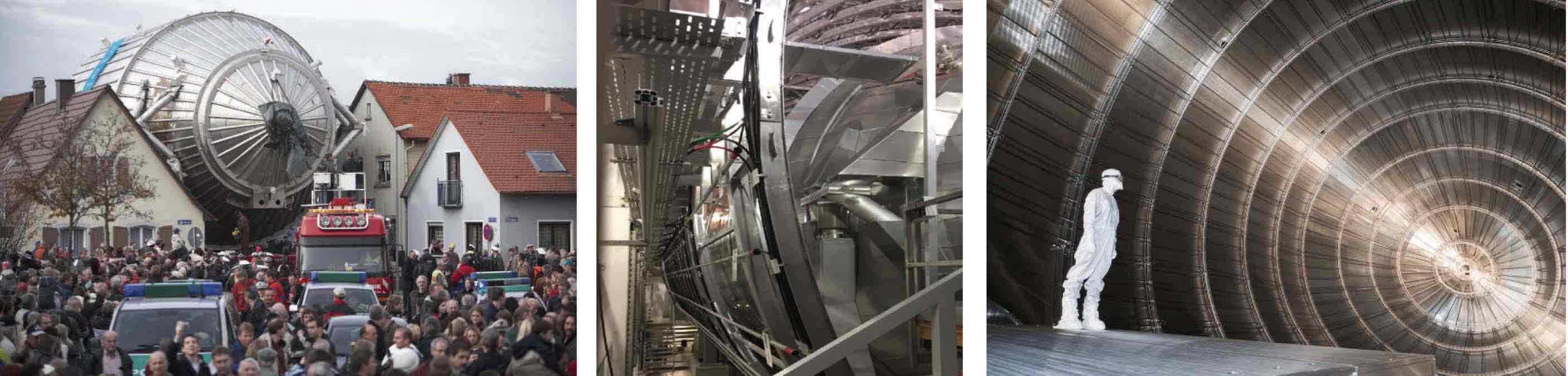}
\caption{The MAC-E vacuum vessel under transport to the Karlsruhe Institute of Technology and two views (one exterior and one
interior) of the spectrometer. }
\label{fig:KATRIN_3photos}
\end{figure}

The KATRIN spectrometer achieves its approximately 1 eV electron energy resolution using a MAC-E  Spectrometer, a spectrometer based on the concept of a 
Magnetic Adiabatic Collimator with
an Electrostatic Filter.\cite{Beamson}  Figure~\ref{fig:KATRIN_3photos} (left photo) shows a view of the MAC-E Spectrometer vacuum vessel as it is being transported to the site of the experiment.   The exterior view (center photo) includes some of the air coils that surround the vessel and fine-tune the magnetic field, including compensation for the Earth's magnetic field.  The interior view (right photo) shows vessel's interior is essentially
empty other than electrostatic guard wires near the inner surface used to reflect backgrounds generated in the walls.

\begin{figure}[htb]
\centering
\includegraphics[height=1.25in]{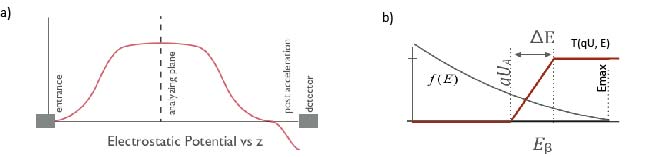}
\caption{(s) Representation of the electrostatic potential energy barrier qU(z) as a function of position in the MAC-E filter.  b) Representations of the
transmission function T(qU,E), shown in red, as a function of electron energy for a fixed retarding potential setting $qU$ and beta energy spectrum $f(E_\beta)$ 
(shown in black) near its endpoint.  }
\label{fig:EFilter}
\end{figure}

To understand the concept of the MAC-E filter, it is perhaps easiest to first consider the electrostatic potential created by KATRIN's electrostatic field and initially ignore the magnetic field.   Letting $z$ denote the longitudinal position with respect to the spectrometer's mid-plane, the electrostatic field, $U(z)$, creates a potential energy barrier,
$qU(z)$, for the electrons shown schematically in Fig.~\ref{fig:EFilter}a.  The value of the potential barrier at the mid-point is known as the {\em retarding energy},
$qU$.   As electrons enter the spectrometer, their kinetic energy can be written as a sum of transverse and longitudinal energies.    Electrons with a longitudinal energy greater than the electrostatic barrier at mid-point, or {\em analyzing plane}, cross the analyzing plane
and are accelerated back to their original energy as they travel to the exit point of the spectrometer.  A post-acceleration region 
can be used to provide an additional 9 keV of kinetic energy;
when this is enabled the electrons originating with energies near the 18.6 keV tritium endpoint reach the detector with about 27 keV of energy and are easily detected in the Focal Plan Detector (FPD) with high efficiency.

The FPD simply counts the number of electrons that reach it.  For each potential setting, the count rate, $R(qU_A)$,  is a function of the height
of the potential energy barrier at the analyzing plane, $qU_A$.   For a fixed retarding energy, the rate is determined by the beta spectrum
$f(E_\beta)$, convoluted by the approximately step-like {\em transmission function}, $T(qU_A,E)$:
$$
R(qU_A) = \int_{qU{_A}}^{E_{max}} T(qU_A,E_\beta) f(E_\beta) dE_\beta.
$$

The red line in Fig~\ref{fig:EFilter}b represents the transmission function and the black line represents the beta  energy spectrum, $f(E_\beta)$,
 near its endpoint. The width of the transition region, $\Delta$E, is the {\em filter width}.   It is determined by the maximum transverse energy of the electrons as they reach the analyzing plane.
If one collimated the electrons down to a narrow beam at the source, their kinetic energy would be nearly 100\% longitudinal energy, giving
an abrupt transition from zero to 100\% transmission;  $\Delta$E would be very small.   Unfortunately, extreme collimation would yield a  low count rate and is not practical.

\begin{figure}[htb]
\centering
\includegraphics[height=1.7in]{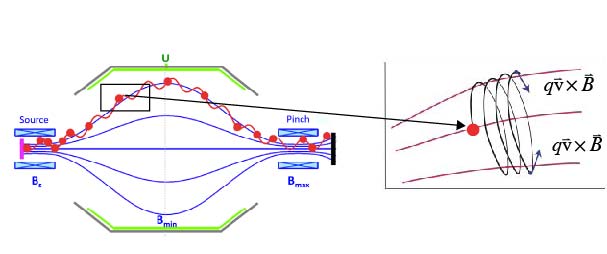}
\caption{The magnetic field lines within the MAC-E filter.  The inset illustrates the cyclotron motion of an electron as it spirals about a field line in a region
of diverging field lines results in a component of the Lorentz force along the central field line, resulting in the transformation of transverse kinetic energy to
longitudinal kinetic energy. }
\label{fig:MLines}
\end{figure}

We now consider the functions of the magnetic field constituting the {\em Magnetic Adiabatic Collimator} part of the MAC-E spectrometer (illustrated in
Fig.~\ref{fig:MLines}) that  provides the
mechanism for minimizing  the transverse electron energy, and thus the filter width, while maintaining a large angular acceptance in order to obtain high statistics. The electrons follow the field lines, spiraling around them as the magnetic field guides the electrons from the spectrometer's entrance to its 
exit. Hence the {\em collimator} part of the MAC-E filter.  The 6~T field at the entrance of the spectrometer drops to $3\times10^{-4}$~T at the analyzing plane, a decrease of more than 4 orders of magnitude.  The inset in Fig.~\ref{fig:MLines} shows that, as an electron spirals about a field line in a region of decreasing field strength, it experiences a Lorentz force which has a small component along the central field line since the field lines are diverging as the field decreases in magnitude.  There is no change in the total kinetic energy, but the transverse kinetic energy is transformed to additional longitudinal kinetic energy.   

 As an electron follows a field line, the change in field strength is small over one cyclotron orbit and thus the {\em adiabatic} approximation can be used to easily quantify
 the  transformation of transverse kinetic energy to longitudinal kinetic energy.  For this situation, the orbital angular magnetic moment is a constant, 
$$
\mu = IA = \left( a\frac{v_\perp}{2\pi r}\right)\left(\pi r^2\right) = \frac{E_\perp}{B} = \textrm{constant}
$$

This results in a transverse kinetic energy reduction that is proportional to the reduction in magnetic field. Including a small relativistic effect,  the filter width, $\Delta E$,
due to the maximum remaining transverse kinetic energy as the electrons reach the analyzing plane can be written as
$$
\frac{\Delta  E}{E} = \frac{B_A}{B_{max}} \frac{\gamma +1}{2}
$$
 The standard KATRIN magnetic configurations of $B_{max}$=6~T and $B_{anal~plane}$=0.3~mT gives a filter width of  0.93 eV for
 electrons near
  the 18.6 keV tritium beta decay endpoint.   The complete model of the response function includes effects from inelastic scattering of the
  betas in the WGTS, Doppler broadening, and modifications to the beta spectrum due to tritium molecular final state binding
  effects.\cite{Kleesiek:2018}

The electrons are guided to the Focal Plane Detector by the magnetic field lines.   The electrostatic retarding potential at the analyzing plane is actually a function of the radial distance from the axis of symmetry, but the electron's position as it passes through the analyzing plane is mapped to a specific pixel on the detector.  We can analyze the data using the appropriate stopping potential value for a given trajectory.   The detector's keV-scale energy resolution allows us to put cuts on a region of interest to reduce background events.

The KATRIN apparatus includes a host of additional instrumentation and capabilities used for calibration and monitoring not addressed
in this paper, but critical to obtaining reliable results.

\section{First Light and Krypton Campaigns}

What we call "First Light" was achieved in October, 2016 when electrons, generated through the photoelectric effect on the rear wall, were transported through the entire apparatus and detected in the pixelated focal plane detector.   This initial running of the KATRIN spectrometer was used to verify calculations of the flux tube diameter, other system parameters.

\begin{figure}[htb]
\centering
\includegraphics[height=1.5in]{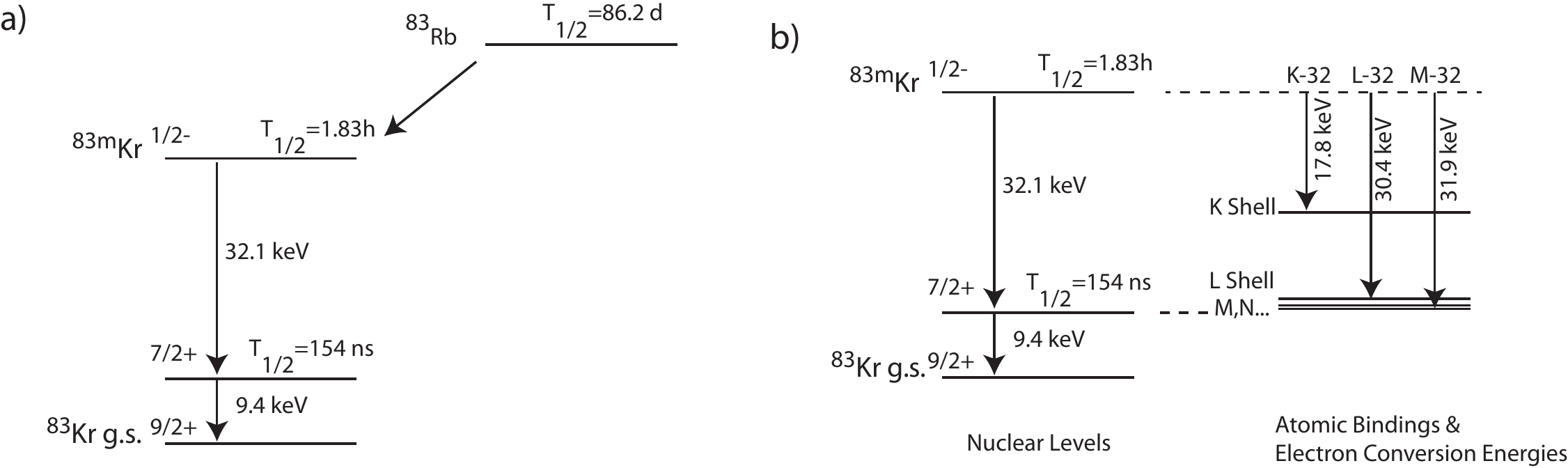}
\caption { a) The relevant energy levels of the $^{83}$Kr  nuclear energy levels fed from the beta decay of $^{83}$Rb.   b) The $^{83}$Kr electron
shell levels and the corresponding electron conversion energies associated with the nuclear 32.1 keV de-excitation.}
\label{fig:KryptonSimplified}
\end{figure}

KATRIN uses a metastable state of krypton, $^{83m}$Kr, as a source of pseudo-mono energetic electrons.     This state, shown in 
Fig.~\ref{fig:KryptonSimplified}a,
 is created from the decays of Rubidium and has a half-life of 1.83 h;  it can be used within the apparatus with no long-term contributions to backgrounds. 
 The metastable state decays through two successive electron conversions.   Figure~\ref{fig:KryptonSimplified}b, shows that the first 32.1~keV nuclear
 de-excitation results in the ejection of a conversion electron whose energy depends on the initial atomic shell of the  electron.   This process results a series
 of pseudo mono-energetic electron lines denoted K-32, L-32, M-32, etc.  with the energies shown in the figure.   The subsequent nuclear de-excitation from
 the 9.4~keV nuclear level also occurs through electron conversion, resulting in a second series of ejected electron energies labeled L-9.4, M-9.4, etc.
 (The 9.4~keV nuclear de-excitation does not provide sufficient energy to eject a K-shell electron.)  The KATRIN spectrometer's  sub-eV resolution is actually
 sensitive to the initial atomic sub-shell energy splittings, so the complete conversion electron labeling convention includes the sub-shell.   For example,  L2-32
 notation denotes the a 32.1~keV nuclear de-excitation ejecting an electron from the second sub-shell of the L-shell atomic krypton atom.

The KATRIN apparatus includes two krypton sources within the apparatus in addition to one used in the HV monitoring spectrometer.
A small condensed krypton source can be inserted  and moved about to test the mapping of source location within the flux tube to specific pixels in the Focal Plane Detector. 
In addition, gaseous krypton can be injected into the WGTS to generate a homogenous distribution in the detector.    This can be done while maintaining tritium in the WGTS and allows measurements of the transmission function, including effects due to the electrons interacting with the tritium gas.

\begin{figure}[htb]
\centering
\includegraphics[height=1.6in]{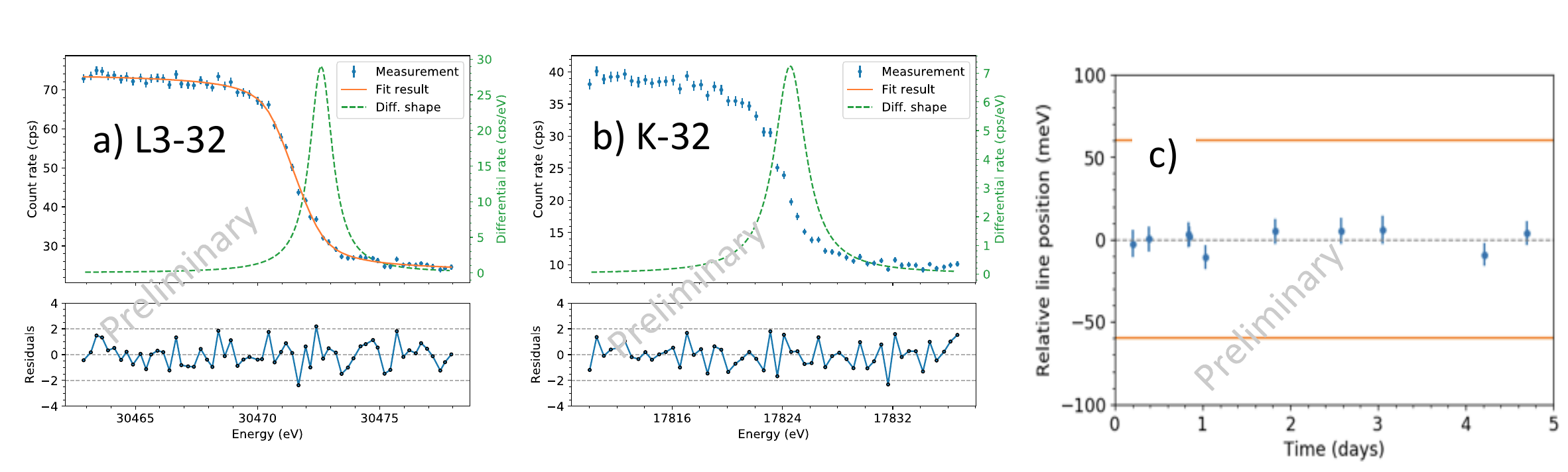}
\caption { Results from the first gaseous krypton measurements. a)  The FPD count rate as a function of retarding energy setting shown as blue
data points.  The orange curve is a 4-parameter fit to the data and the green curve is the corresponding extracted L3-32 line shape.  b) Similar for a scan over the
K-32 line position.  c) The relative extracted line position as a function of time.}
\label{fig:KryptonResults}
\end{figure}

Figure~\ref{fig:KryptonResults} shows the results from a gaseous krypton run performed in the summer of 2017. 
  In Fig.~\ref{fig:KryptonResults}a, the data points
(blue) show the measured FPD counting rates as the retarding energy is stepped through a series of energies, sweeping through the energy
corresponding to  the krypton
L3-32 electron conversion line.   The pseudo-monoenergic conversion electron data are modeled with three parameters:
 line position, width, and amplitude and a fourth parameter is used to allow a flat background under the conversion electron peak.
This 4-parameter characterization is then folded with the calculated KATRIN response function and fit to the data.  The resulting fit is shown as the orange curve and
the corresponding line shape, given by these four fitted parameters, is shown as the dashed green line in Fig.~\ref{fig:KryptonResults}a.  
Figure~\ref{fig:KryptonResults}b shows similar results
for a scan over the K-32 line energy.   The stability of the KATRIN spectrometer is demonstrated in Fig.~\ref{fig:KryptonResults}c, which shows the variation in
the fitted line position as a function of time over five days of running.   This demonstrated stability is well within the KATRIN design limit of $\pm60$meV as shown in the figure.

As of the date of this conference, KATRIN has published two papers on the {\em First Light} and {\em Krypton} campaigns and a half-dozen more
are undergoing the KATRIN Collaboration review process.\cite{Arenz:2018a}\cite{Arenz:2018b}

\section{Status and Outlook}
The very first tritium was injected into the WGTS on May 18, 2018, only a few weeks before this conference.   Initial ion-safety and sub-system check outs were performed
using a D$_2$ carrier gas with 1\% tritium.   The first tritium beta-decay energy spectra were measured the following day.   At this time,
the commissioning is continuing with the first low-concentration tritium running and the initial results are being prepared for the
{\em Neutrino 2018} conference.    Over the next year, we expect to make additional commissioning and
krypton measurements and then proceed to acquire the initial tritium datasets.   We expect to initiate a  search for
kinks in the beta spectrum that could signify the existence of a sterile neutrino in the keV mass range.  This search could be {\em Phase 0}
of a program being developed to search for signatures of sterile neutrinos.

\Acknowledgements
I am  grateful for the opportunity to present the work of the entire KATRIN contribution and, in particular, the contributions of Larisa Thorne and Prof. Diana Parno to the preparation of this talk.    The KATRIN collaboration acknowledges the support of Helmholtz Association (HGF), Ministry for Education and Research BMBF (05A17PM3, 05A17PX3, 05A17VK2, and 05A17WO3), Helmholtz Alliance for Astroparticle Physics (HAP), and Helmholtz Young Investigator Group (VH-NG-1055) in Germany; Ministry of Education, Youth and Sport (CANAM-LM2011019), cooperation with the JINR Dubna (3+3 grants) 2017Ð2019 in the Czech Republic; and the Department of Energy through grants DE-FG02-97ER41020, DE-FG02-94ER40818, DE-SC0004036, DE-FG02-97ER41033, DE-FG02-97ER41041, DE-AC02-05CH11231, DE-SC0011091 and DE-SC0019304 in the United States.

\end{document}